\documentclass{article}
\usepackage{spconf,amsmath,graphicx,amsfonts}
\usepackage{cite}
\usepackage{amsmath}
\usepackage{amssymb}
\usepackage{caption}
\usepackage{enumitem}
\usepackage{amsmath,times}
\usepackage{mathtools,amssymb}

\DeclareMathOperator*{\argmin}{arg\,min}
\title{\Large Study of Channel Estimation with Oversampling for 1-bit Large-Scale MIMO Systems}

\name{Zhichao~Shao, Lukas~T.~N.~Landau and Rodrigo~C.~de~Lamare}
\address{Centre for Telecommunications Studies\\
    Pontifical Catholic University of Rio de Janeiro,
    Rio de Janeiro, Brazil 22453-900\\
    Email: zhichao.shao;lukas.landau;delamare@cetuc.puc-rio.br}

\begin{document}
\maketitle
\begin{abstract}
In this paper, we propose an oversampling based low-resolution aware least squares channel estimator for large-scale multiple-antenna systems with 1-bit analog-to-digital converters on each receive antenna. To mitigate the information loss caused by the coarse quantization, oversampling is applied at the receiver, where the sampling rate is faster than the Nyquist rate. We also characterize analytical performances, in terms of the deterministic Cram\'er-Rao bounds, on estimating the channel parameters. Based on the correlation of the filtered noise, both the Fisher information for white noise and a lower bound of Fisher information for colored noise are provided. Numerical results are provided to illustrate the mean square error performances of the proposed channel estimator and the corresponding Cram\'er-Rao bound as a function of the signal-to-noise ratio.
\end{abstract}

\section{Introduction}
Large-scale multiple-input multiple-output (MIMO) systems are regarded as a promising candidate for the next generation communication systems, as they offer significant increases in data throughput without an additional increase in bandwidth or transmit power \cite{Larsson,mmimo,6457363,lasmismatch}. However, the use of more antennas at the base station (BS) will bring more challenges, such as system complexity and pilot contamination \cite{6375940}. One problem faced is the high energy consumption of the system while using high speed, high precision (e.g. 8-12 bits) analog-to-digital converters (ADCs) at each front-end (RF). A possible solution is to use a number of ultra low cost and ultra low precision ADCs (1-3 bits) \cite{7894211,7458830,6987288,1bitcpm}. These low cost ADCs can largely reduce the energy consumption of the BS. The performance loss caused by the coarse quantization can be partially recovered by using some adequate compensation methods like oversampling. Moreover, the low cost ADCs can also be used in millimeter-Wave (mmWave) systems, which can achieve larger bandwidths of 500 MHz or more. The works in \cite{8337813,7094595,6804238,1bit_oce} have studied the channel estimation, signal detection, achievable rate and precoding techniques in such systems.

Currently, large-scale MIMO systems with 1-bit ADCs at the RF have attracted much attention, as they can largely reduce the receiver energy consumption. Recent studies include precoding \cite{8010806}, channel estimation \cite{8385500}, capacity analysis \cite{7155570} and iterative detection and decoding (IDD) technique \cite{8240730}. To reduce the quantization errors caused by the 1-bit quantizer, oversampling is applied at each receive antenna, where the sampling rate is significantly higher than the Nyquist rate. The authors in \cite{8487039} have studied the situation that sampling is faster than symbol rate (FTSR) in an uplink massive MIMO system with 1-bit ADCs. For acquiring the channel state information (CSI), they develop a Bussgang-based linear minimum mean squared error (BLMMSE) channel estimator for the oversampled system, which is based on impractical noise assumptions and consumes a higher computational cost.



In this paper, we propose a low-resolution aware least squares (LRA-LS) channel estimator for uplink massive MIMO systems with 1-bit quantization and oversampling at the receiver based on the Bussgang theorem. In particular, we develop an adaptive recursion to estimate the autocorrelation of the channel vector required in the expression of the LRA-LS estimator, which can also be used in the BLMMSE channel estimator. For non-oversampled systems, we analyze the Fisher information (FI) and present the Cramér-Rao Bound (CRB) for any unbiased estimator. In contrast to the noise assumption in \cite{8487039}, where it is assumed that the filtered noise samples are uncorrelated, we consider correlated noise samples, which is  important for oversampled systems. Since the exact expression of the FI is still an open problem, we present lower bounds of the FI.

The rest of the paper is organized as follows. The system model is
shown in section 2. In section 3, we illustrate the FI for both
non-oversampled and oversampled 1-bit MIMO systems. In section 4, we
give a short derivation of the proposed oversampling based LRA-LS
channel estimator. In section 5, the simulation results are
presented and we conclude the paper in section 6.

The following notations are used throughout the paper. $a$ is a scalar, $\mathbf{a}$ is a vector, $\mathbf{A}$ is a matrix. The $n\times n$ identity matrix is denoted by $\mathbf{I}_n$ and $\mathbf{0}_n$ is a $n\times 1$ all zeros column vector. We use $\text{diag}(\mathbf{A})$ to denote a diagonal matrix only containing the diagonal elements of $\mathbf{A}$. $\mathbf{A}^T$, $\mathbf{A}^H$, $\mathbf{A}^*$ and $\mathbf{A}^+$ represent transpose, conjugate transpose, complex conjugate and pseudo-inverse, respectively. $\otimes$ denotes the Kronecker product. $[\mathbf{a}]_k$ gets the $k$th element from $\mathbf{a}$. $[\cdot]^R$ and $[\cdot]^I$ represents the real and imaginary part, respectively.

\begin{figure*}[!htbp]
    \centering
        \includegraphics[width=19.4cm, height=6.3cm]{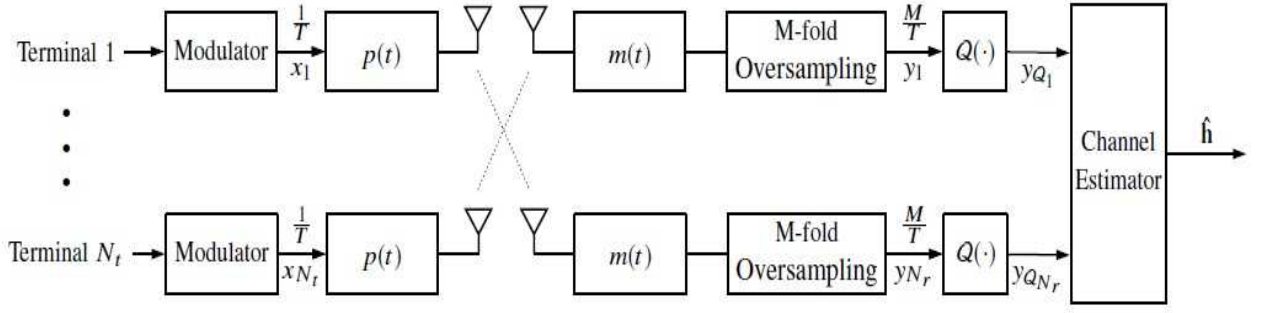}
    \caption{System model of 1-bit multi-user multiple-antenna system with oversampling at the receiver}
    \label{fig:transmitter}
\end{figure*}

\section{System Model}
Consider the uplink of a single-cell multi-user large-scale MIMO system, as shown in Fig. 1, where there are $N_t$ single-antenna terminals and the BS is equipped with $N_r$ receive antennas. For the large-scale MIMO system we have $N_r\gg N_t$. The received oversampled signal $\mathbf{y}$ is
\begin{equation}
\mathbf{y}=\mathbf{H}\mathbf{x}+\mathbf{n},
\label{equ:system_model}
\end{equation}
where $\mathbf{x}\in\mathbb{C}^{NN_t \times 1}$ contains $N$ blocks of transmitted symbols. Each symbol is generated by independent and identically distributed (i.i.d.) random variable with zero mean and unit variance. The vector $\mathbf{n}\in\mathbb{C}^{MN_rN \times 1}$ contains filtered oversampled noise samples expressed by
\begin{equation}
\mathbf{n} = \left(\mathbf{I}_{N_r}\otimes \mathbf{G}\right)\mathbf{w}.
\end{equation}
The entries of $\mathbf{w}$ are zero-mean i.i.d. complex Gaussian random variables, which we
denote by $\mathcal{CN}\left(\mathbf{0}_{3MN_rN},\sigma^2_n\mathbf{I}_{3MN_rN}\right)$. Since the receive filter length is $2MN + 1$, more samples (e.g. $3MN$) in $\mathbf{w}$ need to be considered to have the same statistical property of each entry in $\mathbf{n}$. $\mathbf{G}$ is a Toeplitz matrix described by (\ref{equ:R}), which contains the coefficients of the matched filter $m(t)$ at different time instants,
\begin{figure*}[!htbp]
    \begin{equation}
    \mathbf{G} = \begin{bmatrix}
    m(-NT)& m(-NT+\frac{1}{M}T)& \dots& m(NT) & 0 & \dots & 0\\
    0 & m(-NT)& \dots & m(NT-\frac{1}{M}T) & m(NT) & \dots & 0\\
    \vdots & \vdots & \ddots & \vdots & \vdots & \ddots & \vdots\\
    0 & 0 & \dots & m(-NT)& m(-NT+\frac{1}{M}T)& \dots& m(NT)\\
    \end{bmatrix}_{MN\times 3MN}
    \label{equ:R}
    \end{equation}
\end{figure*}
where the symbol period is represented by $T$.

The equivalent channel matrix $\mathbf{H}\in\mathbb{C}^{MN_rN \times NN_t}$ is described as
\begin{equation}
\mathbf{H} = \left(\mathbf{I}_{N_r}\otimes \mathbf{Z}\right)\mathbf{U}\left(\mathbf{H}'\otimes \mathbf{I}_N\right),
\label{equ:H}
\end{equation}
where $\mathbf{H}'\in\mathbb{C}^{N_r \times N_t}$ is the channel matrix for the non-oversampled systems and $\mathbf{U}$ is the oversampling matrix described by
\begin{equation} 
\mathbf{U} = \mathbf{I}_{N_rN}\otimes\mathbf{u} = \mathbf{I}_{N_rN}\otimes \left[0 \quad \cdots \quad 0 \quad 1\right]^T_{1 \times M}.
\end{equation}
$\mathbf{Z}\in\mathbb{R}^{MN \times MN}$ is a Toeplitz matrix with the form
\begin{equation}
\resizebox{.46\textwidth}{!}{$\displaystyle
    \mathbf{Z} = \begin{bmatrix}
    z(0) & z(\frac{T}{M}) & \dots & z(NT-\frac{1}{M}T)\\
    z(-\frac{T}{M}) & z(0) & \dots & z(NT-\frac{2}{M}T)\\
    \vdots & \vdots & \ddots & \vdots\\
    z(-NT+\frac{1}{M}T) & z(-NT+\frac{2}{M}T) & \dots & z(0)\\
    \end{bmatrix}$},
\end{equation}
where $z(t)$ is the convolution of the pulse shaping filter $p(t)$ and the matched filter $m(t)$. In particular, $M=1$ refers to the non-oversampling case.

Let $\mathcal{Q}(.)$ represent the 1-bit quantization function, the received quantized signal $\mathbf{y}_\mathcal{Q}$ is
\begin{equation}
\mathbf{y}_\mathcal{Q}=\frac{1}{\sqrt{2}}\mathcal{Q}\left(\mathbf{y}\right)=\frac{1}{\sqrt{2}}\mathcal{Q}\left(\mathbf{y}^R\right) + j\frac{1}{\sqrt{2}}\mathcal{Q}\left(\mathbf{y}^I\right),
\label{system_model}
\end{equation}
The real and imaginary parts are element-wisely quantized to $\{\pm1\}$ based on the sign, respectively.


\section{Fisher Information and Channel Estimation for 1-bit MIMO}
This section firstly analyzes the FI about the unknown channel parameters and gives deterministic CRBs on the variance of any unbiased channel estimator for both non-oversampled and oversampled 1-bit MIMO systems. The proposed LRA-LS channel estimator and the adaptive estimation of the autocorrelation matrix of the channel vector is then described in the last part.

The system model in (\ref{equ:system_model}) can be vectorized as
\begin{equation}
\begin{aligned}
\mathbf{y}&=(\mathbf{x}^T\otimes\mathbf{I}_{N_rNM})\text{vec}(\mathbf{H})+\mathbf{n}\\&=[\mathbf{x}^T\otimes\mathbf{I}_{N_r}\otimes\mathbf{Z}(\mathbf{I}_{N}\otimes\mathbf{u})]\text{vec}(\mathbf{H}'\otimes \mathbf{I}_N)+\mathbf{n}
\end{aligned}
\label{equ_sys}
\end{equation}
with the property of vectorization and Kronecker products
\begin{equation}
\begin{aligned}
\text{vec}(\mathbf{H}'\otimes \mathbf{I}_N)&=\\&\hspace{-1cm} \left[\mathbf{I}_{N_t}\otimes\left(\mathbf{e}_1\otimes\mathbf{I}_{N_r}\otimes\mathbf{e}_1+\dots+\mathbf{e}_N\otimes\mathbf{I}_{N_r}\otimes\mathbf{e}_N\right)\right]\text{vec}(\mathbf{H'}),
\end{aligned}
\end{equation}
where $\mathbf{e}_n$ is an all zeros column vector except that the $n$th element is one. Eq.(\ref{equ_sys}) can be written in the following simplified form
\begin{equation}
\mathbf{y}=\mathbf{\Phi}\text{vec}(\mathbf{H'})+\mathbf{n}=\mathbf{\Phi}\mathbf{h}'+\mathbf{n},
\label{Equ_nonquan}
\end{equation}
where $\mathbf{\Phi}\in\mathbb{C}^{MNN_r \times N_rN_t}$ is called the equivalent transmit matrix.

To calculate the FI, we rewrite (\ref{Equ_nonquan}) in the real-valued form as
\begin{equation}
\begin{bmatrix}
\mathbf{y}^R \\
\mathbf{y}^I
\end{bmatrix}=\begin{bmatrix}
\mathbf{\Phi}^R &-\mathbf{\Phi}^I\\
\mathbf{\Phi}^I & \mathbf{\Phi}^R
\end{bmatrix}\begin{bmatrix}
\mathbf{h'}^R \\
\mathbf{h'}^I
\end{bmatrix}+\begin{bmatrix}
\mathbf{n}^R \\
\mathbf{n}^I
\end{bmatrix}.
\label{equ:system_model_real}
\end{equation}
Considering the unknown parameter vector $\tilde{\mathbf{h'}}=[\mathbf{h'}^R;\mathbf{h'}^I]$ and with the independence of $\mathbf{y}^R$ and $\mathbf{y}^I$, the FI matrix \cite{Kay:1993:FSS:151045} of the quantized signal is defined as
\begin{equation}
\mathbf{F}_{\mathbf{y}_\mathcal{Q}}(\tilde{\mathbf{h'}})=\mathbf{F}_{\mathbf{y}_\mathcal{Q}^R}(\tilde{\mathbf{h'}})+\mathbf{F}_{\mathbf{y}_\mathcal{Q}^I}(\tilde{\mathbf{h'}}),
\end{equation}
where
\begin{equation}
    [\mathbf{F}_{\mathbf{y}_\mathcal{Q}^{R/I}}(\tilde{\mathbf{h'}})]_{ij}=E_{\mathbf{y}_\mathcal{Q}^{R/I}\mid\tilde{\mathbf{h'}}}\left\{\frac{\partial \ln p(\mathbf{y}_\mathcal{Q}^{R/I}\mid\tilde{\mathbf{h'}})}{\partial [\tilde{\mathbf{h'}}]_i}\frac{\partial \ln p(\mathbf{y}_\mathcal{Q}^{R/I}\mid\tilde{\mathbf{h'}})}{\partial [\tilde{\mathbf{h'}}]_j}\right\}
\end{equation}
with $[\tilde{\mathbf{h'}}]_i$ and $[\tilde{\mathbf{h'}}]_j$ being the elements of $\tilde{\mathbf{h'}}$.
The variance of any unbiased estimator $\hat{\tilde{\mathbf{h'}}}$ is lower bounded by
\begin{equation}
    \text{var}\{[\hat{\tilde{\mathbf{h'}}}]_i\}\geq[\mathbf{F}_{\mathbf{y}_\mathcal{Q}}^{-1}(\tilde{\mathbf{h'}})]_{ii}.
\end{equation}

\subsection{Fisher Information for Non-oversampled Systems}
For non-oversampled system, i.e, $M=1$, the noise vector $\mathbf{n}$ contains white Gaussian noise samples with covariance matrix $\mathbf{C}_\mathbf{n}=\sigma_n^2\mathbf{I}_{NN_r}$. The log-likelihood function can be expressed as
\begin{equation}
\ln p(\mathbf{y}_\mathcal{Q}\mid\tilde{\mathbf{h'}})=\sum_{k=1}^{NN_r}\left[\ln p([\mathbf{y}_\mathcal{Q}^R]_k\mid\tilde{\mathbf{h'}})+\ln p([\mathbf{y}_\mathcal{Q}^I]_k\mid\tilde{\mathbf{h'}})\right]
\end{equation}
with
\begin{equation}
p\left([\mathbf{y}_\mathcal{Q}^{R}]_k=\frac{1}{\sqrt{2}}\mid\tilde{\mathbf{h'}}\right)=Q\left(-\frac{[\mathbf{\Phi}^R\mathbf{h'}^R-\mathbf{\Phi}^I\mathbf{h'}^I]_k}{\sigma_n/\sqrt{2}}\right)
\end{equation}
\begin{equation}
p\left([\mathbf{y}_\mathcal{Q}^{R}]_k=-\frac{1}{\sqrt{2}}\mid\tilde{\mathbf{h'}}\right)=Q\left(\frac{[\mathbf{\Phi}^R\mathbf{h'}^R-\mathbf{\Phi}^I\mathbf{h'}^I]_k}{\sigma_n/\sqrt{2}}\right)
\end{equation}
\begin{equation}
p\left([\mathbf{y}_\mathcal{Q}^{I}]_k=\frac{1}{\sqrt{2}}\mid\tilde{\mathbf{h'}}\right)=Q\left(-\frac{[\mathbf{\Phi}^I\mathbf{h'}^R+\mathbf{\Phi}^R\mathbf{h'}^I]_k}{\sigma_n/\sqrt{2}}\right)
\end{equation}
\begin{equation}
p\left([\mathbf{y}_\mathcal{Q}^{I}]_k=-\frac{1}{\sqrt{2}}\mid\tilde{\mathbf{h'}}\right)=Q\left(\frac{[\mathbf{\Phi}^I\mathbf{h'}^R+\mathbf{\Phi}^R\mathbf{h'}^I]_k}{\sigma_n/\sqrt{2}}\right),
\end{equation}
where $Q(x) = \frac{1}{\sqrt{2\pi}}\int_x^\infty \exp(-\frac{u^2}{2})du$. With the derivative of $Q(x)$ function, the FI for the real part is given by
\begin{equation}
\begin{aligned}
[\mathbf{F}_{\mathbf{y}_\mathcal{Q}}^R(\tilde{\mathbf{h'}})]_{ij} &= \sum_{k=1}^{NN_r}-E\left\{\frac{\partial^2\ln p([\mathbf{y}_\mathcal{Q}^R]_k\mid\tilde{\mathbf{h'}})}{\partial [\tilde{\mathbf{h'}}]_i\partial [\tilde{\mathbf{h'}}]_j}\right\}\\&\hspace{-1.7cm}=\frac{1}{\pi\sigma_n^2}\sum_{k=1}^{NN_r}\frac{\exp(-\frac{[\mathbf{\Phi}^R\mathbf{h'}^R-\mathbf{\Phi}^I\mathbf{h'}^I]_k^2}{\sigma_n^2/2})\frac{\partial[\mathbf{\Phi}^R\mathbf{h'}^R-\mathbf{\Phi}^I\mathbf{h'}^I]_k}{\partial [\tilde{\mathbf{h'}}]_i}\frac{\partial[\mathbf{\Phi}^R\mathbf{h'}^R-\mathbf{\Phi}^I\mathbf{h'}^I]_k}{\partial [\tilde{\mathbf{h'}}]_j}}{Q\left(\frac{[\mathbf{\Phi}^R\mathbf{h'}^R-\mathbf{\Phi}^I\mathbf{h'}^I]_k}{\sigma_n/\sqrt{2}}\right)Q\left(-\frac{[\mathbf{\Phi}^R\mathbf{h'}^R-\mathbf{\Phi}^I\mathbf{h'}^I]_k}{\sigma_n/\sqrt{2}}\right)}.
\end{aligned}
\end{equation}
The derivation for the imaginary part is analogous.

\subsection{Fisher Information for Oversampled Systems}
When $M\geq2$ the equivalent noise vector $\mathbf{n}$ contains correlated noise samples. Computing the exact form of $p(\mathbf{y}_\mathcal{Q}^{R/I}\mid\tilde{\mathbf{h'}})$ is not available or it is too difficult to compute. Instead, the authors in \cite{6783980} have given a lower bound of the FI, which is based on the first and second order moments
\begin{equation}
    \mathbf{F}_{\mathbf{y}_\mathcal{Q}^{R/I}}(\tilde{\mathbf{h'}})\geq\left(\frac{\partial \mathbf{\mu}_{\mathbf{y}_\mathcal{Q}^{R/I}}}{\partial \tilde{\mathbf{h'}}}\right)^T\mathbf{C}^{-1}_{\mathbf{y}^{R/I}_\mathcal{Q}}\left(\frac{\partial \mathbf{\mu}_{\mathbf{y}_\mathcal{Q}^{R/I}}}{\partial \tilde{\mathbf{h'}}}\right) = \tilde{\mathbf{F}}_{\mathbf{y}_\mathcal{Q}^{R/I}}(\tilde{\mathbf{h'}}),
\end{equation}
where the equality holds for $M=1$. Since the lower-bounding technique is identical for the real and the imaginary part, we present only the derivation of $\tilde{\mathbf{F}}_{\mathbf{y}_\mathcal{Q}^R}(\tilde{\mathbf{h'}})$. Based on \cite{8445905}, the mean value of the $k$th received symbol is given by
\begin{equation}
\begin{aligned}
[\mathbf{\mu}_{\mathbf{y}_\mathcal{Q}^R}]_k &= \frac{1}{\sqrt{2}}P\left([\mathbf{y}_\mathcal{Q}^R]_k=+1\mid\tilde{\mathbf{h'}}\right)-\frac{1}{\sqrt{2}}P\left([\mathbf{y}_\mathcal{Q}^R]_k=-1\mid\tilde{\mathbf{h'}}\right)\\&=\frac{1}{\sqrt{2}}\left[1-2Q\left(\frac{[\mathbf{\Phi}^R\mathbf{h'}^R-\mathbf{\Phi}^I\mathbf{h'}^I]_k}{\sqrt{[\mathbf{C}_\mathbf{n}]_{kk}/2}}\right)\right],
\end{aligned}
\label{equ_miu}
\end{equation}
The derivative of (\ref{equ_miu}) is
\begin{equation}
\frac{\partial [\mathbf{\mu}_{\mathbf{y}_\mathcal{Q}^R}]_k}{\partial [\tilde{\mathbf{h'}}]_i}=\frac{2\text{exp}\left(-\frac{[\mathbf{\Phi}^R\mathbf{h'}^R-\mathbf{\Phi}^I\mathbf{h'}^I]_k^2}{[\mathbf{C}_\mathbf{n}]_{kk}}\right)\frac{\partial [\mathbf{\Phi}^R\mathbf{h'}^R-\mathbf{\Phi}^I\mathbf{h'}^I]_k}{\partial [\tilde{\mathbf{h'}}]_i}}{\sqrt{2\pi[\mathbf{C}_\mathbf{n}]_{kk}}}.
\end{equation}
The diagonal elements of the covariance matrix are given by
\begin{equation}
[\mathbf{C}_{\mathbf{y}^R_\mathcal{Q}}]_{kk}=\frac{1}{2}-[\mathbf{\mu}_{\mathbf{y}_\mathcal{Q}^R}]_k^2,
\end{equation}
while the off-diagonal elements are calculated as
\begin{equation}
\begin{aligned}
[\mathbf{C}_{\mathbf{y}^R_\mathcal{Q}}]_{kn}=&P(z_k>0,z_n>0)+P(z_k\leq0,z_n\leq0)\\&-\frac{1}{2}-[\mathbf{\mu}_{\mathbf{y}_\mathcal{Q}^R}]_k[\mathbf{\mu}_{\mathbf{y}_\mathcal{Q}^R}]_n,
\end{aligned}
\end{equation}
where $[z_k,z_n]^T$ is a bi-variate Gaussian random vector
\begin{equation*}
\begin{bmatrix}
z_k\\z_n
\end{bmatrix}\sim\mathcal{N}\left(\begin{bmatrix}
[\mathbf{\Phi}^R\mathbf{h'}^R-\mathbf{\Phi}^I\mathbf{h'}^I]_k\\ [\mathbf{\Phi}^R\mathbf{h'}^R-\mathbf{\Phi}^I\mathbf{h'}^I]_n
\end{bmatrix},\frac{1}{2}\begin{bmatrix}
[\mathbf{C}_\mathbf{n}]_{kk} & [\mathbf{C}_\mathbf{n}]_{kn}\\
[\mathbf{C}_\mathbf{n}]_{nk} & [\mathbf{C}_\mathbf{n}]_{nn}
\end{bmatrix}\right).
\end{equation*}
The lower bound for the imaginary part is derived in the same way.

\subsection{Oversampling based LRA-LS Channel Estimation}
In each uplink transmission block pilots are located before the data symbols. During the training phase, all terminals simultaneously transmit $\tau$ pilot sequences to the BS. Eq.(\ref{Equ_nonquan}) yields
\begin{equation}
\mathbf{y}_p=\mathbf{\Phi}_p\mathbf{h'}+\mathbf{n}_p.
\label{equ_sysmod}
\end{equation}
In particular, for the 1-bit quantization and with the Bussgang theorem (\ref{equ_sysmod}) can be decomposed as
\begin{equation}
    \mathbf{y}_{\mathcal{Q}_p}= \mathcal{Q}(\mathbf{\Phi}_p\mathbf{h'} + \mathbf{n}_p)=\tilde{\mathbf{\Phi}}_p\mathbf{h'} + \tilde{\mathbf{n}}_p,
    \label{equ_linear}
\end{equation}
where $\tilde{\mathbf{\Phi}}_p\in\mathbb{C}^{M\tau N_r \times N_tN_r} = \mathbf{A}_p\mathbf{\Phi}_p$ and $\tilde{\mathbf{n}}_p\in\mathbb{C}^{M\tau N_r \times 1} = \mathbf{A}_p\mathbf{n}_p+\mathbf{n}_q$. The vector $\mathbf{n}_q$ is the statistically equivalent quantizer noise. The matrix $\mathbf{A}_p\in\mathbb{R}^{M\tau N_r \times M\tau N_r}$ is the linear operator chosen independently from $\mathbf{y}_p$, which yields
\begin{equation}
\mathbf{A}_p=\mathbf{C}_{\mathbf{y}_p\mathbf{y}_{\mathcal{Q}_p}}^H\mathbf{C}_{\mathbf{y}_p}^{-1}=\sqrt{\frac{2}{\pi}}\text{diag}\left(\mathbf{C}_{\mathbf{y}_p}\right)^{-\frac{1}{2}},
\end{equation}
where $\mathbf{C}_{\mathbf{y}_p\mathbf{y}_{\mathcal{Q}_p}}$ denotes the cross-correlation matrix between the received signal $\mathbf{y}_p$ and the quantized signal $\mathbf{y}_{\mathcal{Q}_p}$ \cite{Bussgang}
\begin{equation}
\mathbf{C}_{\mathbf{y}_p\mathbf{y}_{\mathcal{Q}_p}}=\sqrt{\frac{2}{\pi}}\text{diag}(\mathbf{C}_{\mathbf{y}_p})^{-\frac{1}{2}}\mathbf{C}_{\mathbf{y}_p}.
\end{equation}
$\mathbf{C}_{\mathbf{y}_p}$ is the auto-correlation matrix of $\mathbf{y}_p$, as follows:
\begin{equation}
\mathbf{C}_{\mathbf{y}_p}=\mathbf{\Phi}_p\mathbf{R}_\mathbf{h'}\mathbf{\Phi}_p^H+\sigma_n^2\left(\mathbf{I}_{N_r}\otimes \mathbf{GG}^H\right).
\label{equ_Cy}
\end{equation}
Based on the equivalent linear model (\ref{equ_linear}), the LS estimate of $\mathbf{h'}$ is given by
\begin{equation}
\begin{aligned}
\hat{\mathbf{h'}}_{LS}&=\argmin_{\bar{\mathbf{h'}}}\quad (\mathbf{y}_{\mathcal{Q}_p}-\tilde{\mathbf{\Phi}}_p\bar{\mathbf{h'}})^H(\mathbf{y}_{\mathcal{Q}_p}-\tilde{\mathbf{\Phi}}_p\bar{\mathbf{h'}})\\&=(\tilde{\mathbf{\Phi}}_p^H\tilde{\mathbf{\Phi}}_p)^{-1}\tilde{\mathbf{\Phi}}_p^H\mathbf{y}_{\mathcal{Q}_p}.
\end{aligned}
\label{equ_LS}
\end{equation}


In practice, $\mathbf{R}_\mathbf{h'}=E\{\mathbf{h'}\mathbf{h'}^H\}$ is not known at the receiver, which is a limitation for the BLMMSE channel estimator. We propose an adaptive technique to recursively estimate it as
\begin{equation}
\hat{\mathbf{R}}_\mathbf{h'}[n+1] = \lambda\hat{\mathbf{R}}_\mathbf{h'}[n] + \hat{\mathbf{h'}}[n]\hat{\mathbf{h'}}[n]^H,\quad 1\leq n\leq \tau,
\end{equation}
where $\lambda$ is the forgetting factor and $\hat{\mathbf{h'}}[n]$ is the channel estimate at time instant n. Consider the system model
\begin{equation}
\begin{aligned}
\mathbf{y}_\mathcal{Q}[n]&=\mathcal{Q}(\mathbf{H}\mathbf{x}[n]+\mathbf{n}[n])\\&=\mathcal{Q}\left((\mathbf{x}^T[n]\otimes\mathbf{I}_{N_r}\otimes\mathbf{Z'u})\mathbf{h'}+\mathbf{n}[n]\right),
\end{aligned}
\end{equation}
where $\mathbf{y}_\mathcal{Q}[n]$, $\mathbf{x}[n]$ and $\mathbf{n}[n]$ are column vectors with size $MN_r\times1$, $N_t\times1$ and $MN_r\times1$, respectively. $\mathbf{Z'}\in\mathbb{R}^{M \times M}$ is a simplified version of $\mathbf{Z}$. The instantaneous estimate of $\mathbf{h'}$ is calculated as
\begin{equation}
\hat{\mathbf{h'}}[n] =  \left(\mathbf{x}^T[n]\otimes\mathbf{I}_{N_r}\otimes\mathbf{Z'u}\right)^+\mathbf{y}_\mathcal{Q}[n].
\end{equation}
The initial guess of $\hat{\mathbf{R}}_\mathbf{h'}[1]$ is an all-zeros matrix. Note that the proposed LRA-LS channel estimator is a biased estimator. While calculating the Cram\'er-Rao upper bound, it should apply as follows
\begin{equation}
\frac{\partial E\{\hat{\tilde{\mathbf{h'}}}_{LS}\}}{\partial \tilde{\mathbf{h'}}}\left(\tilde{\mathbf{F}}_{\mathbf{y}_\mathcal{Q}}^{-1}(\tilde{\mathbf{h'}})\frac{\partial E\{\hat{\tilde{\mathbf{h'}}}_{LS}\}}{\partial \tilde{\mathbf{h'}}}\right)^T.
\end{equation}
Instead of directly calculating the gradient of expectation with respect to the channel vector, we numerically evaluate this gradient, since there is an adaptive estimation technique inside the channel estimator. Other estimators \cite{jio,jidf,jiomimo} can also be considered.

\section{Numerical Results}

\begin{figure}[!htbp]
    \centering
    \includegraphics[width=8.4cm, height=6.3cm]{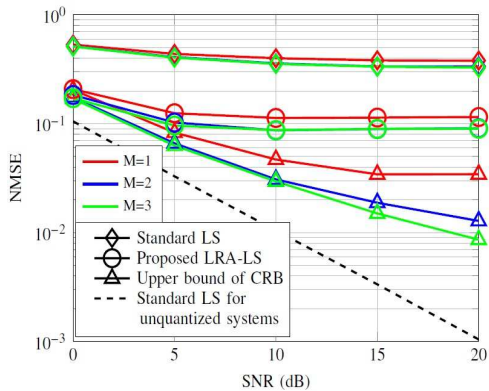}
    \caption{$N_t = 4$ and $N_r = 16$. NMSE comparison between different oversampling factors with $\tau = 40$.}
    \label{fig:MSE}
\end{figure}

\begin{figure}[!htbp]
    \centering
    \includegraphics[width=8.4cm, height=6.3cm]{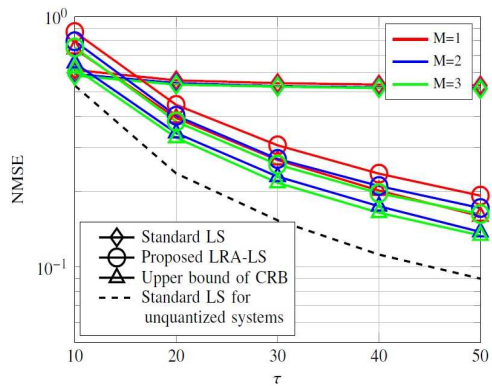}
    \caption{$N_t = 4$ and $N_r = 16$. NMSE comparison as a function of pilot length when $\text{SNR}=0\text{dB}$.}
    \label{fig:MSE_pilots}
\end{figure}

This section presents simulation results of the proposed
oversampling based LRA-LS channel estimation algorithm. We have used
QPSK as the modulation scheme. The $m(t)$ and $p(t)$ filters are
normalized Root-Raised-Cosine (RRC) filters with a roll-off factor
of 0.8. The channel is assumed to experience block fading and
$\mathbf{C}_\mathbf{h'}=\mathbf{I}_{N_rN_t}$. The forgetting factor
$\lambda$ is set to 0.91. The pilot sequences are orthogonal. The
signal-to-noise ratio (SNR) is defined as
$10\log(\frac{N_t}{\sigma_n^2})$. Fig. \ref{fig:MSE} shows the
normalized mean square error (NMSE) performance of the proposed
channel estimator and the analytical upper bound of CRB. When using
the LRA-LS channel estimator, there is a 2 dB performance gain of
the oversampled system ($M=2$ or 3) compared to the non-oversampled
system ($M=1$). The upper bound of CRB is calculated based on the
corresponding lower bound of FI. Fig. \ref{fig:MSE_pilots} shows the
NMSE performances of the system with different lengths of pilot
symbols at SNR=0dB, where the proposed channel estimator approaches
the performance of the upper bound of CRB.

In the simulations we have chosen $\tau=40$ as the trade-off between
system complexity and estimation performance. The symbol error rate
(SER) performances of the oversampled system with estimated CSI
(channel state information) and perfect CSI are shown in Fig.
\ref{fig:MSE_pilots}, where the sliding window based linear receiver
with window length 3 \cite{8450809} is applied in the system. Other
detection schemes \cite{spa,mfsic,mbsic,dfcc,mbdf} can also be
considered.

\section{Conclusion}
This work has proposed the LRA-LS channel estimator for uplink
large-scale MIMO systems with 1-bit quantization and oversampling at
the receiver. We have further given analytical performance of the
system in terms of the FI. For the non-oversampled system, the
equivalent noise vector $\mathbf{n}$ contains white noise samples
and the lower bound of FI is the same as the actual FI. The
simulation results have shown that the proposed channel estimator in
oversampled systems achieves better performance than that of
non-oversampled systems. $\mathcal{F}$

\section*{\bfseries{Acknowledgements}}
This work has been supported in part by the ELIOT ANR-18-CE40-0030 and FAPESP 2018/12579-7 project
and the CNPq Universal 438043/2018-9 project.

\bibliographystyle{IEEEbib}
\bibliography{ref}

\end{document}